\documentclass[intlimits,twoside,a4paper]{article}

\usepackage{amsmath,amssymb}
\usepackage{graphicx}

\usepackage[T2A]{fontenc}
\usepackage[cp1251]{inputenc}

\usepackage{cmpj2}

\usepackage{multirow}
\usepackage{booktabs}

\issue{2013}{16}{3}{33602}
\doinumber{10.5488/CMP.16.33602}

\title[Phase transformation B1 to B2 in TiC, TiN, ZrC and ZrN under pressure]%
	{Phase transformation B1 to B2 in TiC, TiN, ZrC and ZrN under pressure}
\author[V.I. Ivashchenko, P.E.A. Turchi, V.I. Shevchenko]%
	{V.I. Ivashchenko\refaddr{label1}, P.E.A. Turchi\refaddr{label2},
    V.I. Shevchenko\refaddr{label1}}
\addresses{
	\addr{label1} Institute of Problems of Material Science of the
		National Academy of Sciences of Ukraine, \\
3 Krzhyzhanovsky St., 03142 Kyiv, Ukraine
	\addr{label2} Lawrence Livermore National Laboratory (L--352),
		P.O. Box 808, Livermore, CA 94551, USA
}

\date{Received December 14, 2012, in final form June 6, 2013}
\authorcopyright{V.I. Ivashchenko, P.E.A. Turchi, V.I. Shevchenko, 2013}

\begin{document}

\maketitle

\begin{abstract}
	Phase stability of various phases of MX (M\,=\,Ti, Zr; X\,=\,C, N) at equilibrium and under 	 pressure is examined based on first-principles calculations of the electronic and phonon 	 structures.	The results reveal that all B1 (NaCl-type) MX structures undergo a phase transition to the B2-structures under high pressure in agreement with the previous total-energy calculations.	The B1-MX structures are dynamically stable under very high pressure ($210\div570$~GPa).	The pressure-induced B2 (CsCl-type) MC phases are dynamically unstable even at high pressures, and	TiN and ZrN are found to crystallize with the B2-structure only at pressures above $55$~GPa.	The first-order B1-to-B2 phase transition in these nitrides is not related to the softening of	phonon modes, and the dynamical instability of B2-MX is associated with a high density of states at the Fermi level.
\pacs 64.70.K-, 63.20.-e, 71.15.Nc, 71.20.-b
\end{abstract}

\section{Introduction}

Transition metal compounds (TMC) form a class of materials with the NaCl-type crystal structure (B1) in a wide composition range, and exhibit extremely high melting points, hardness, and metallic conductivity \cite{1,2}.
These materials are exploited under extreme conditions (high temperature, high pressure,
{etc.}), and they are widely used as the main layers in ultra-hard nano-composite coatings \cite{3}. Therefore, it is important to understand the peculiarities of their behavior under extreme conditions. In this investigation we focus on titanium and zirconium carbides and nitrides (MX, M\,=\,Ti, Zr; X\,=\,C, N) that crystallize with the NaCl-type (B1) lattice since their high-pressure stability is not very well known.
To our knowledge, there has been only several works devoted to this problem: Dubrovinskaia et al. observed a B1-R phase transition in TiC at a pressure above 18~GPa \cite{4}, although Winkler et al. found no phase transition in the B1 titanium carbide up to 26~GPa \cite{5}. On the other hand, many theoretical studies predict the B1 to B2 (CsCl-type) phase transition in MX, under high pressure.
In table~\ref{tab:1}, we report the transition pressures ($P_0$) obtained in the present
investigation and from other studies based on first-principles and empirical-potential calculations. It is seen that all the first-principles and three-body potential calculations provide consistent values of $P_0$, whereas the neglect of three body interactions in the empirical-potential studies leads to much lower values of $P_0$ (cf.~table~\ref{tab:1}).
Based on the total-energy calculations for TiC, Zhao {et al.} predicted two phase transition paths under pressure: B1 $\rightarrow$ (117~GPa) TlI$'$ $\rightarrow$ (172~GPa) TlI, and/or B1 $\rightarrow$ (131~GPa) TiB$'$ $\rightarrow$ (148~GPa) TiB \cite{6}.
Note that a description of these phases can be found in reference~\cite{6}.

It follows from this brief review that many theoretical investigations predict a B1 to B2 phase transition in MX under pressure in the range of $200\div550$~GPa.
Also, the more complex cubic to orthorhombic transformations under pressure in TiC and ZrC were predicted in some first-principles calculations.
For MX, there is no experimental evidence of the existence of other structural polytypes,
besides the B1 structure, in a wide range of temperatures (up to melting points) and pressures.

The main goal of this paper is to verify the dynamical stability of the B2-MX phases under pressure, and to investigate other plausible phases of MX.

The paper is organized as follows.
In section~\ref{sec:2}, we present our theoretical framework and the computational details.
Section~\ref{sec:3} contains the results of our calculations together with comments.
Finally, section~\ref{sec:4} contains the main conclusions.

\begin{table}[!t]
\caption{Transition pressures, $P_0$ (in GPa), associated with the B1 $\rightarrow$ B2 phase transformation for MX according to different electronic structure methods. }
\label{tab:1}
	\vspace{2ex}
	\begin{center}
		\begin{tabular}{|l|l|l|l|}
			\hline
			Phase &	$P_0$ & Procedure & Reference \\
			\hline\hline
			\multirow{4}*{TiC}		&	570     & Quantum ESPRESSO-GGA & 	This work \\
		      	&	500     & CASTEP-GGA           & 	\cite{6} \\
			      &	490     & FPLMTO-LDA           & 	\cite{7} \\
			      &	57      & Two-body potential   & 	\cite{8} \\
			\hline
			\multirow{6}*{TiN}		&	354     & Quantum ESPRESSO-GGA & 	This work \\
			      &	370     & FPLMTO-LDA           & 	\cite{7} \\
			      & 364.1   & CASTEP-GGA           & 	\cite{9} \\
			      & 322.2   & CASTEP-LDA           & 	\cite{9} \\
			      & 126     & Two-body potential   & 	\cite{8,10} \\
			      & 310     & Three-body potential & 	\cite{11} \\
			\hline
			\multirow{3}*{ZrC}  	& 289     & Quantum ESPRESSO-GGA & 	This work \\
			      & 295     & CASTEP-GGA           & 	\cite{12} \\
			      & 98      & Two-body potential   & 	\cite{13} \\
			\hline
			\multirow{3}*{ZrN}	  & 209     & Quantum ESPRESSO-GGA & 	This work \\
			      & 205     & CASTEP-GGA           & 	\cite{12} \\
			      & 94      & Two-body potential   & 	\cite{10} \\
			\hline
		\end{tabular}
	\end{center}
\end{table}

\section{Methodology\label{sec:2}}

A first-principles pseudo-potential method was used to investigate different phases
of MX (M\,=\,Ti, Zr; X\,=\,C, N).
Scalar-relativistic band-structure calculations within the density functional theory (DFT) were carried out.
The ``Quantum-ESPRESSO'' first-principles code \cite{14} was used to perform the pseudo-potential
calculations, with Vanderbilt ultra-soft pseudo-potentials to describe electron-ion interactions \cite{15}.
For Ti and Zr atoms, the semi-core states were treated as valence states.
Plane waves up to a kinetic energy cutoff of 38~Ry (516.8~eV) were included in the basis set. The exchange-correlation potential was treated in the framework of the generalized gradient approximation (GGA) of Perdew-Burke-Ernzerhof (PBE) \cite{16}.
Brillouin-zone integrations were performed using the sets of special $k$-points corresponding to the ($8\times8\times8$) Monkhorst-Park mesh \cite{17}.
For the 8-atom cell of the TlI-type TiC, we used a ($8\times4\times8$) mesh that, although it
generates a lower number of $k$-points, provides an acceptable accuracy.
Each eigenvalue was convoluted with a Gaussian with width $\sigma = 0.02$~Ry (0.272~eV).
All structures were optimized by simultaneously relaxing the atomic basis vectors and the atomic
positions inside the unit cells using the Broyden-Fletcher-Goldfarb-Shanno (BFGS) algorithm
\cite{18}.
The relaxation of the atomic coordinates and the unit cell was considered to be complete when the
atomic forces were less than 1.0~mRy/Bohr (25.7~meV/{\AA}), the stresses were smaller than
0.025~GPa, and the total energy during the structural optimization iterative process varied by
less than 0.1~mRy (1.36~meV).
The crystalline and energetic parameters of various phases of titanium and zirconium carbides and
nitrides after structural optimization are summarized in table~\ref{tab:2}.
The electronic densities of states (DOS) were calculated using the ($12\times12\times12$) mesh.

The above-described pseudo-potential procedure was used to study the phonon spectra of the NaCl (B1) and CsCl (B2)-type MX compounds in the framework of the density-functional perturbation theory (DFPT) described in references~\cite{14,19}.
The first-principles DFPT calculations were carried out for a ($8\times8\times8$) $q$-mesh, and then the phonon densities of states (PHDOS) were computed using a ($12\times12\times12$) $q$-mesh by interpolating the computed phonon dispersion curves.
Both the DOS and PHDOS were calculated using the tetrahedron method implemented in the
``Quantum-ESPRESSO'' code \cite{14}.

\begin{table}[t]
	\vspace*{-6mm}
	\caption{Symmetry, number of atoms per unit cell ($N_\textrm{a}$), structural parameters, $a$,
		$b$, $c$ (in \AA), and atomic volume, $V$ (in \AA$^3$/atom), and total energy, $E_\textrm{T}$ (in
		eV/atom), of the different phases (with associated struktur-bericht designation) of TiC, TiN, ZrC
		and ZrN at equilibrium.
		Calculated structural parameters from other experimental (in parentheses) and theoretical (in
		curly brackets) investigations are also reported.}
	\label{tab:2}
	\vspace{2ex}
	\begin{center}
		\begin{tabular}{@{}|l*{9}{|l}|@{}}
			\hline
			Phase & Type & Space Group & No & $N_\textrm{a}$ & $a$ & $b$ & $c$ & $V$ & $E_\textrm{T}$ \\
\hline\hline
			\multirow{10}*{TiC} &
			$\delta$-TiC (B1) & Fm\=3m & 225 & 2 & 4.318 & 4.318 & 4.318 & 10.064 & $-781.554$ \\
			& & & & & (4.318)$^\textrm{a}$   & & & & \\
			& & & & & (4.327)$^\textrm{b}$   & & & & \\
			& & & & & \{4.327\}$^\textrm{c}$ & & & & \\
			& AsNi (B8) & P6$_3$/mmc & 194 & 4 & 3.170 & 3.170 & 4.608 & 10.025 & $-871.398$ \\
			& TlI & Cmcm & 63 & 8 & 2.953 & 8.951 & 3.112 & 10.259 & $-871.120$ \\
			& NiAs (B8) & P6$_3$/mmc & 194 & 4 & 3.032 & 3.032 & 5.356 & 10.660 & $-871.075$ \\
			& WC (B$_\textrm{h}$)& P\=6m2 & 187 & 2 & 3.054 & 3.054 & 2.656 & 10.727 & $-870.888$ \\
			& CoSn (B35)& P6/mmm & 191 & 6 & 4.705 & 4.705 & 3.280 & 10.480 & $-870.834$ \\
			& CsCl (B2)& Pm\=3m & 221 & 2 & 2.700 & 2.700 & 2.700 & 9.842 & $-870.402$ \\
			\hline
			\multirow{7}*{TiN}	&
			$\delta$-TiN (B1) & Pm\=3m & 225 & 2 & 4.318 & 4.318 & 4.318 & 10.064 & $-871.554$ \\
			& & & & & (4.318)$^\textrm{a}$   & & & & \\
			& & & & & (4.327)$^\textrm{b}$   & & & & \\
			& & & & & \{4.315\}$^\textrm{c}$ & & & & \\
			& ZnS (B3) & F\=43m & 216 & 2 & 4.595 & 4.595 & 4.595 & 12.127 & $-929.965$ \\
			& WC (B$_\textrm{h}$)& P\=6m2 & 187 & 2 & 2.899 & 2.899 & 2.696 & 9.811 & $-929.928$ \\
			& CsCl (B2) & Pm\=3m & 221 & 2 & 2.636 & 2.636 & 2.636 & 9.158 & $-929.368$ \\
			\hline
			\multirow{6}*{ZrC} &
			$\delta$-ZrC & Fm\=3m & 225 & 2 & 4.706 & 4.706 & 4.706 & 13.028 & $-752.590$ \\
			& & & & & (4.694)$^\textrm{f}$   & & & & \\
			& & & & & (4.730)$^\textrm{g}$   & & & & \\
			& & & & & \{4.705\}$^\textrm{h}$ & & & & \\
			& CsCl (B2) & Pm\=3m & 221 & 2 & 2.946 & 2.946 & 2.946 & 12.784 & $-751.382$ \\
			& & & & & \{2.919\}$^\textrm{h}$ & & & & \\
			\hline
			\multirow{6}*{ZrN} &
			$\delta$-ZrN (B1) & Pm\=3m & 225 & 2 & 4.592 & 4.592 & 4.592 & 12.104 & $-811.290$\\
			& & & & & (4.585)$^\textrm{i}$   & & & & \\
			& & & & & (4.600)$^\textrm{j}$   & & & & \\
			& & & & & \{4.591\}$^\textrm{h}$ & & & & \\
			& CsCl (B2) & Pm\=3m & 221 & 2 & 2.840 & 2.840 & 2.840 & 11.453 & $-810.393$ \\
			& & & & & \{2.836\}$^\textrm{h}$ & & & & \\
			\hline
		\end{tabular}
		\hfil
		\parbox{0.45\textwidth}{\rule{0pt}{4mm}%
			$^\textrm{a}$ X-ray powder diffraction file [089-3828] \\
			$^\textrm{b}$ X-ray powder diffraction file [065-0242] \\
			$^\textrm{c}$ Ref.~\cite{7} \\
			$^\textrm{d}$ X-ray powder diffraction file [038-1420] \\
			$^\textrm{e}$ X-ray powder diffraction file [065-0414]
		}
		\hfil
		\parbox{0.45\textwidth}{\rule{0pt}{4mm}%
			$^\textrm{f}$ X-ray powder diffraction file [065-0332] \\
			$^\textrm{g}$ X-ray powder diffraction file [065-0962] \\
			$^\textrm{h}$ Ref.~\cite{12} \\
			$^\textrm{i}$ X-ray powder diffraction file [078-1420] \\
			$^\textrm{j}$ X-ray powder diffraction file [065-0961]
		}
		\hfil
	\end{center}
\end{table}

To check the acceptability of the chosen conditions for the calculations we estimate the heat of formation of B1-MX, $H_\textrm{f}$, using the expression $H_\textrm{f} = E_{\textrm{T}} - \Sigma\, n_i E_i$, where $E_{\textrm{T}}$ is the total energy of the bulk compound with $n_i$ atoms of all the involved elements $i$ (Ti, Zr, C, and N), and $E_i$ is the total energy of the bulk hexagonal close-packed structure for Ti or Zr (space group P6$_3$/mmc, No.~194) and diamond (A4) for C, and half the total energy of N$_2$ molecule for N, respectively. The total energy and equilibrium bond length of N$_2$ molecule were computed using the extended two-atom cubic cell. The bond length of N$_2$ molecule was in agreement with the experiment (1.098~{\AA}) within 1\%. The computed values of $H_\textrm{f}$ for NaCl-type (B1) TiC, TiN, ZrC and ZrN are summarized in table~\ref{tab:3} in comparison with the corresponding experimental and theoretical values from
other studies. We note that although theoretical formation energies are fairly consistent with each other, they are all somewhat higher than the experimentally determined values.
One plausible reason for such a discrepancy for zirconium carbide and nitride was discussed in reference~\cite{29}. The reliability of the DFPT calculations of TMC was proved in reference~\cite{30}. Here, we only compare the calculated phonon spectrum with the experimental dispersion curves for TiN \cite{31} to confirm that the DFPT results correctly reproduce the anomalies in the phonon spectra for TMC.

\begin{table}[!t]
\caption{The calculated heat of formation, $H_\textrm{f}$ (in eV/formula unit), for TiC, TiN, ZrC, and ZrN based on the B1 (NaCl-type) structure with the corresponding values determined from experiment and from other calculations for comparison.}
\label{tab:3}
	\vspace{2ex}
	\begin{center}
		\begin{tabular}{|l|l|l|l|}
			\hline
			Phase	&	$H_\textrm{f}$ & Methodology or Experiment & Reference \\
			\hline\hline
		  \multirow{4}*{TiC} &	$-1.800$   & Quantum ESPRESSO-GGA &	This work \\
						&	$-1.896$                 & Experiment           &	\cite{2}  \\
						&	$-1.359\div-1.973$       & Experiment           &	\cite{20} \\
						&	$-1.780$                 & CASTEP-GGA           &	\cite{21} \\
			\hline
			\multirow{7}*{TiN} &	$-3.487$ & Quantum ESPRESSO-GGA & This work \\
						&	$-3.479$ & Experiment           & \cite{2}  \\
						& $-3.46$  & Experiment           & \cite{22} \\
						& $-3.485$ & Experiment           & \cite{23} \\
						& $-3.92$  & CASTEP-GGA           & \cite{21} \\
						& $-3.43$  & Dmol-GGA             & \cite{24} \\
						& $-3.56$  & FLAPW-GGA            & \cite{25} \\
			\hline
			\multirow{6}*{ZrC} & $-1.846$ & Quantum ESPRESSO-GGA & This work \\
						& $-1.912$ & Experiment           & \cite{2}  \\
						& $-2.08$  & Experiment           & \cite{26} \\
						& $-1.72$  & VASP-GGA             & \cite{27} \\
						& $-1.82$  & FPLMTO-LDA           & \cite{28} \\
						& $-1.64$  & VASP-GGA             & \cite{29} \\
			\hline
			\multirow{3}*{ZrN} & $-3.519$ & Quantum ESPRESSO-GGA &	This work \\
						& $-3.771$ & Experiment           &	\cite{2}  \\
						& $-3.784$ & Experiment           &	\cite{23} \\
			\hline
		\end{tabular}
	\end{center}
\end{table}

\section{Results and discussion\label{sec:3}}

To predict the possible stable structures of MX, at first we performed total energy calculations for different phases of TiC and TiN that were identified for other binary compounds (NbN, WC, MoC, CoSn,
CsCl, NiAs, ZnS, {etc.}) at equilibrium \cite{1,2}.
The structural and energetic characteristics of the most stable phases of TiC and TiN are summarized in table~\ref{tab:2}.
It is worth noting that B1 is  the most stable among all the calculated structures, and a good agreement exists between the computed and experimental structural parameters shown in
table~\ref{tab:2}.

\looseness=-1In order to predict the possible stable phases of TiC and TiN under high pressure we calculated the
total energies ($E_\textrm{T}$) of all the TiC and TiN phases presented in table~\ref{tab:2} as
functions of cell volume ($V$).
The analysis of the calculated volume dependence of the total energies, $E_\textrm{T}(V)$, enabled us
to identify only the CsCl-type structures that could be derived from the B1-phases at high pressure.
Under these circumstances, for ZrC and ZrN, we performed total-energy calculations for only two phases
with the NaCl (B1) and CsCl (B2)-type structures (cf.~table~\ref{tab:2}).
We did not find any stable orthorhombic phase of TiC at pressure, as it was predicted in
reference~\cite{6}: the TiB-type TiC phase automatically transformed to the B1 structure during static
relaxation, and the B1-type TiC did not transform into the TlI-type TiC under pressure, although the
latter structure was quite stable at equilibrium (cf.~table~\ref{tab:2}).
Such a discrepancy in the prediction of new phases in TiC is attributed to an inappropriate
procedure of geometry optimization used in reference~\cite{6}.

\begin{figure}[!t]
		\centerline{
			\includegraphics[width=0.4\textwidth]{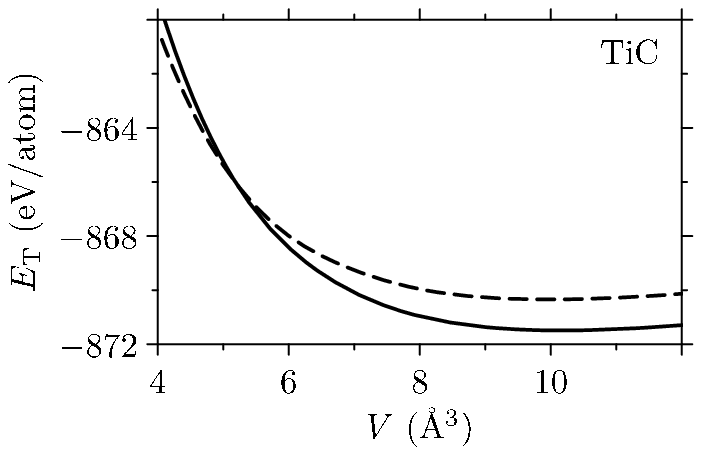}
			\hspace{1mm}
			\includegraphics[width=0.4\textwidth]{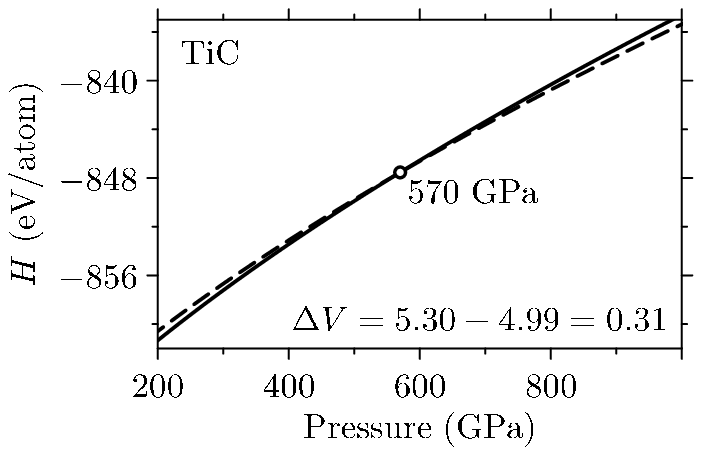}
		}
\vspace{1mm}
		\centerline{
			\includegraphics[width=0.4\textwidth]{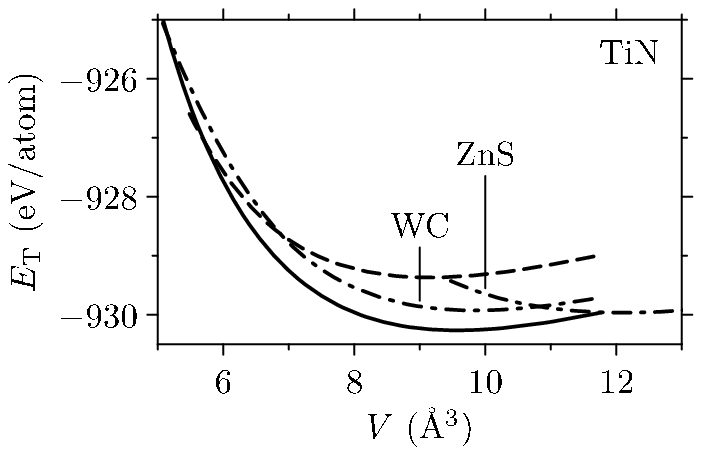}
			\hspace{1mm}
			\includegraphics[width=0.4\textwidth]{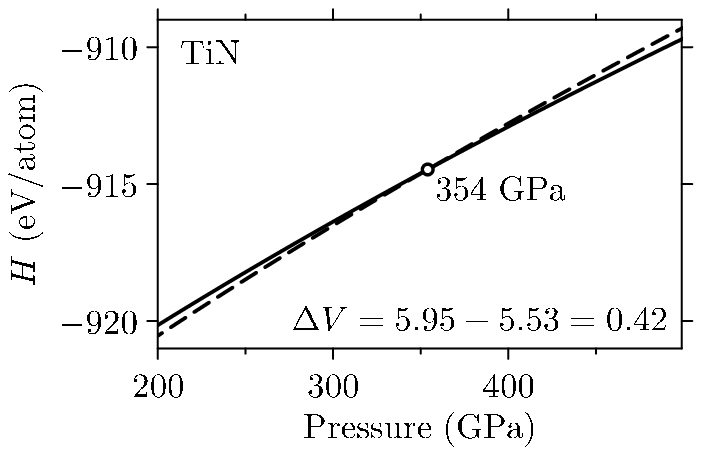}
		}
\vspace{1mm}
		\centerline{
			\includegraphics[width=0.4\textwidth]{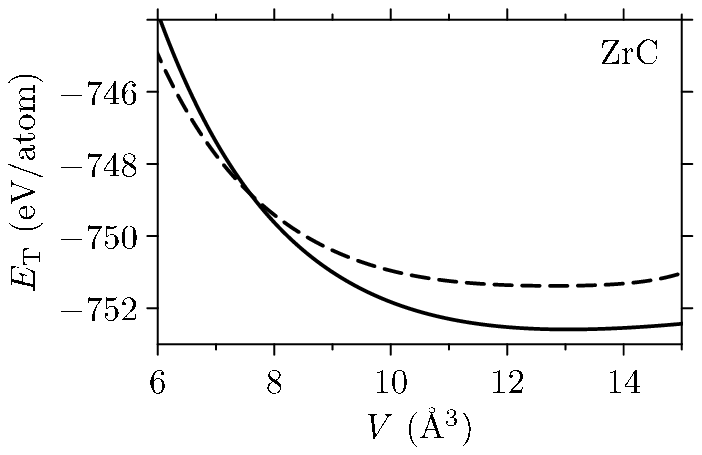}
			\hspace{1mm}
			\includegraphics[width=0.4\textwidth]{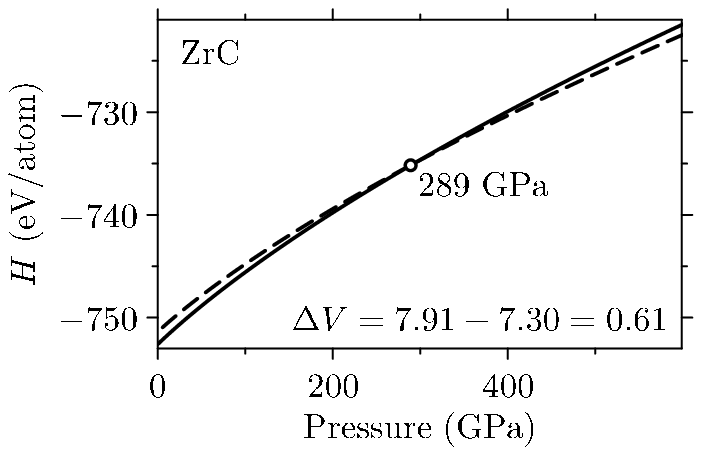}
		}
\vspace{1mm}
		\centerline{
			\includegraphics[width=0.4\textwidth]{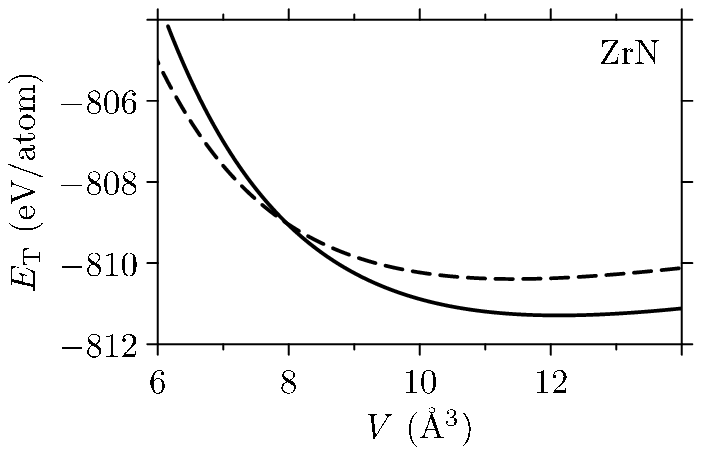}
			\hspace{1mm}
			\includegraphics[width=0.4\textwidth]{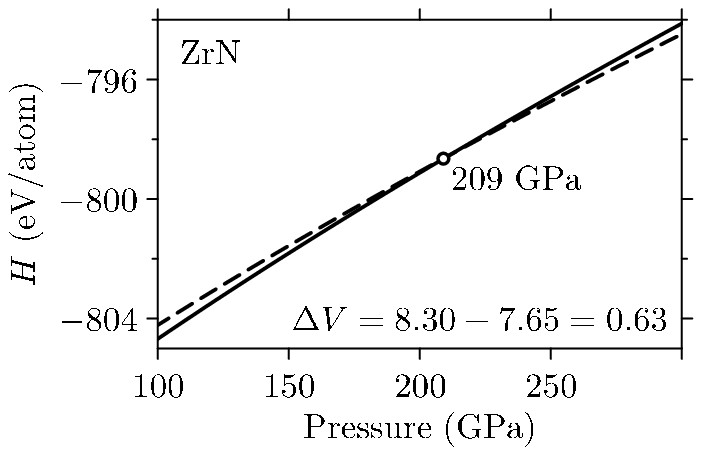}
		}
\caption{Total energy ($E_\text{T}$) as a function of cell volume ($V$), and enthalpy ($H$)
		as a function of pressure for the NaCl (B1)-type (sold line) and CsCl (B2)-type (dashed line)
		phases of TiC, TiN, ZrC and ZrN.
		$E_\textrm{T}(V)$ and $H(P)$ curves are the result of a six-order polynomial fit to the data
		points calculated from the first-principles procedure.
		For TiN, $E_\textrm{T}(V)$ for the WC and ZnS-type structures are also reported.
		$\Delta V$ (in \AA$^3$/atom) is the volume jump at the transition pressure.}
	\label{fig:1}
\end{figure}

The total energies as functions of cell volume and enthalpies as functions of pressure for the B1- and B2-phases are shown in figure~\ref{fig:1}.
One can see that, for the MX phases, the B1-to-B2 first-order phase transition takes place at high
pressure.
The calculated values of transition pressure ($P_0$) are presented in table~\ref{tab:1}, where, for
comparison, the values of $P_0$ obtained from other first-principles and empirical-potential
calculations are also shown.
It is seen that all the calculated characteristics agree well except for $P_0$ calculated from the
two-body empirical potential approach \cite{8,10,13}.
The lower values of $P_0$ in this approach are likely to be a consequence of the neglect of the
many-body interactions.

\begin{figure}[!t]
\centerline{
\includegraphics[width=0.95\textwidth]{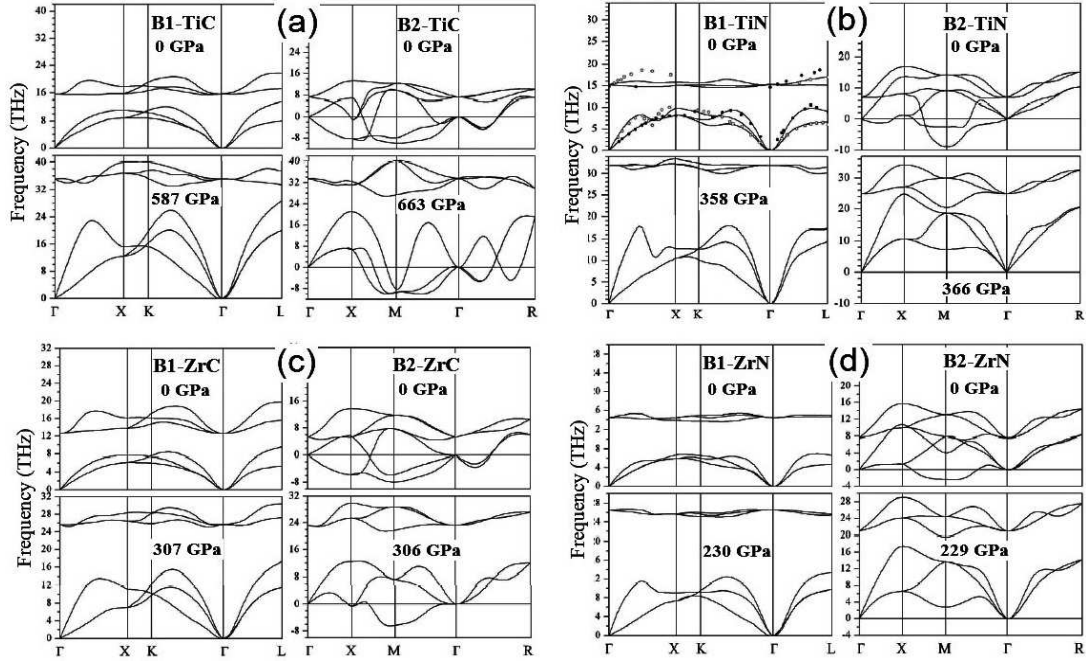}
}
\caption{Phonon dispersion curves along some high-symmetry directions of the BZ for the NaCl (B1) and CsCl (B2)-type of (a) TiC, (b) TiN (The points are the experimental results from
reference~\cite{31}), (c) ZrC, and (d) ZrN at equilibrium and under pressure.
Note that in some portions of the phonon spectra the ``negative'' frequencies are ``imaginary'' (i.\,e., negative squared frequencies).}
\label{fig:2}
\end{figure}

\begin{figure}[!b]
\centerline{
\includegraphics[width=0.5\textwidth]{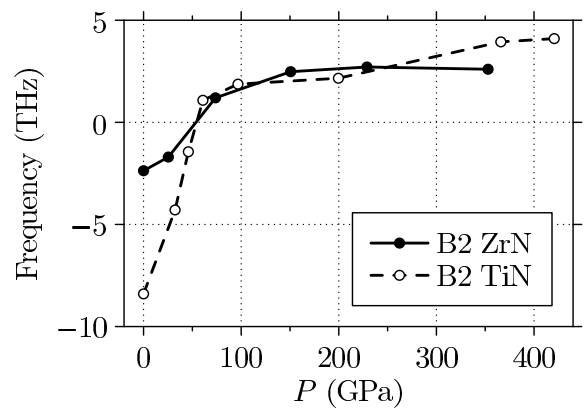}
}
\caption{Frequency of the soft acoustic M$_4$ mode at the M point as a function of pressure  for the CsCl (B2)-type TiN and ZrN.}
\label{fig:3}
\end{figure}

With the B1-to-B2 phase transitions in MX under pressure now being established, we need to
verify that the new pressure-induced phases are dynamically stable.
To address this issue, we calculated the phonon dispersion curves along some high-symmetry
directions of $k$-space for the B1- and B2-phases of MX at equilibrium and under pressure
$P \geqslant P_0$.
The calculated phonon spectra are shown in figure~\ref{fig:2}.
We note that the phonon spectra of all B1-phases do not contain any soft modes, which means that
theses phases should be dynamically stable up to very high pressures.
On the contrary, a softening of the acoustic phonon modes around the X and M points is observed in
the phonon spectrum of the B2-structures at equilibrium, which implies that these phases are
dynamically unstable.
For TiC and ZrC, the condensed phonon modes are preserved even at high pressures
[cf.~figures~\ref{fig:2}~(a),~(c)].
On the other hand, the titanium and zirconium nitrides with the B2 (CsCl-type) structure can be
stabilized by the application of high pressure.
In figure~\ref{fig:3} we show the dependence of the soft M4-modes on pressure for TiN and ZrN. Clearly, the phonon anomalies in these nitrides disappear at $P > 55$~GPa.

\begin{figure}[!t]
\centerline{
\includegraphics[width=0.4\textwidth]{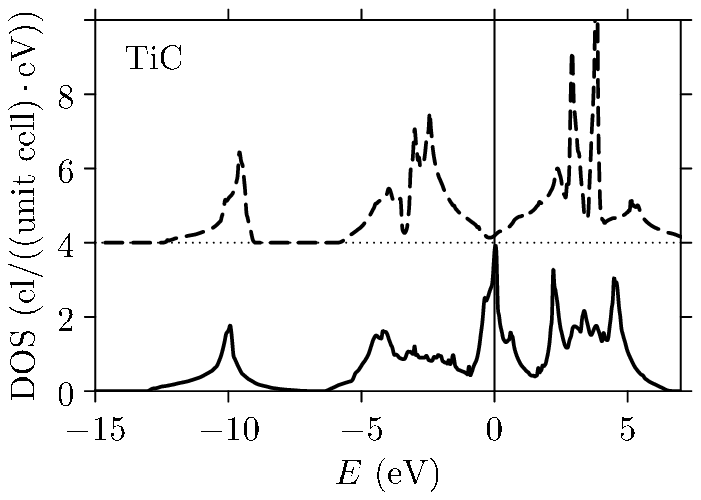}
\hspace{1mm}
\includegraphics[width=0.4\textwidth]{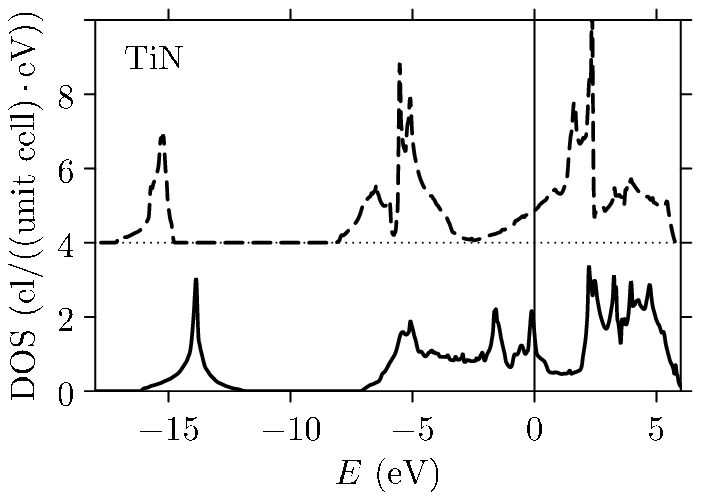}
}
\centerline{
\includegraphics[width=0.4\textwidth]{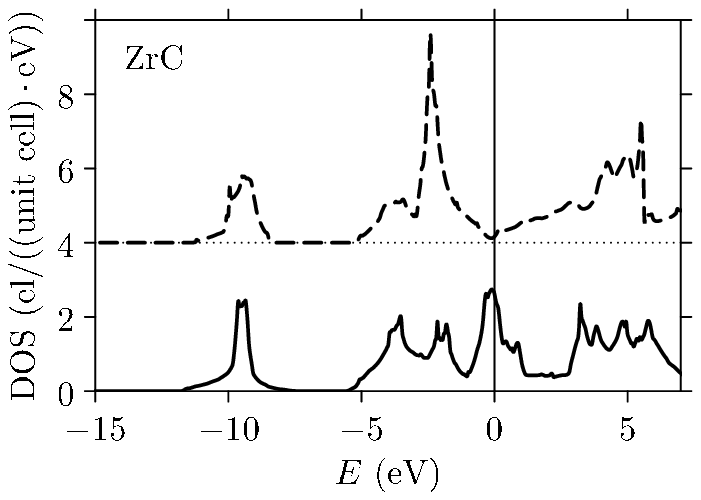}
\hspace{1mm}
\includegraphics[width=0.4\textwidth]{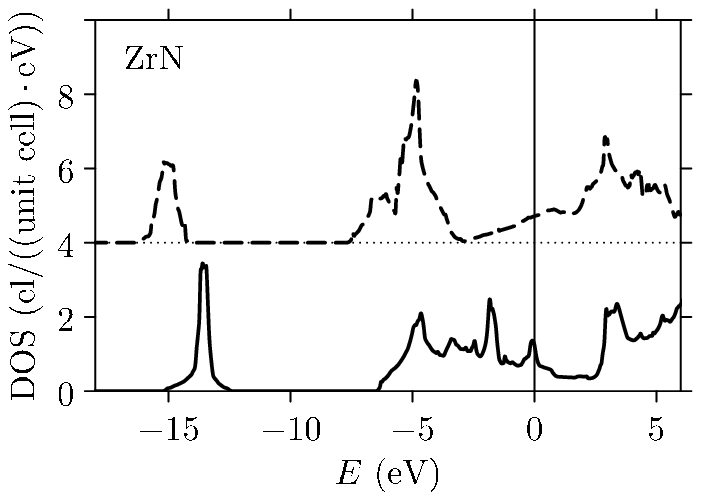}
}
\caption{Densities of states (DOS) of the CsCl (B2) (solid line) and NaCl (B1) (dashed line) type phases of titanium and zirconium carbides and nitrides. The vertical line corresponds to the Fermi level, taken as zero of energy.}
\label{fig:4}
\end{figure}

To understand the plausible origin of the dynamical instability of the B2-MX phases, let us
investigate the electronic structure of the B1 and pressure-induced B2 phases.
The densities of states (DOS) of the B1- and B2-MX phases at equilibrium are shown in
figure~\ref{fig:4}.

In the case of B1-MX, at increasing energy, the lowest bands are associated with the $2s$ states of
X, then the next bands originate from X $2p$- and M $3d$-states, and finally, the broad M $d$-bands
with a small admixture of X $2p$ states are located above the minimum of the DOS.
There is a new type of M-$d_y$ --- M-$d_y$ interactions in B2-based structures that are absent in B1-MX. As a result, several local peaks appear at the bottom of the metal band.
One can see that, for all the B2-MX phases, the Fermi level ($E_\textrm{F}$) crosses the local DOS maximum in
the region of the spectrum formed by the M-$d_y$ states (the partial DOS is not shown here).
A high DOS at the Fermi level is usually associated with a structural instability and with the existence of soft phonon modes in the long-wave region, and the collapse of these modes leads to a structural
transformation.
Thus, the high DOS at the Fermi level identified in the case of B2-MX can be one of the reasons for
the dynamical instability of these phases.

As a final note, since modern high-pressure devices can generate pressures up to 550~GPa that are
higher than the maximum pressure of the earth's core (360~GPa) \cite{32}, we hope that our theoretical
findings will motivate further high-pressure experiments to establish new pressure-induced phases in
transition metal compounds including those that were examined in this work.

\section{Conclusions\label{sec:4}}

Phase stability of various phases of MX (M\,=\,Ti, Zr; X\,=\,C, N) at equilibrium and under pressure
was examined based on the first-principles calculations of the electronic and phonon structures.
The calculated formation energies are in good agreement with the corresponding experimental values.
The analysis of the dependencies of enthalpy and phonon spectra on pressure for these phases enabled
us to bring the following conclusions.

It follows from the total energy calculations that all B1-MX structures undergo a phase
transformation to the B2-structures under high pressure in agreement with the previous total-energy
calculations.
The B1-MX structures are dynamically stable under very high pressure ($209\div570$~GPa).
The calculated phonon spectra show that the B2-MX compounds have got collapsed acoustic modes at ambient
and high pressures.
However, B2-based TiN and ZrN can be dynamically stabilized at pressures above 55~GPa.
The first-order B1-to-B2 phase transition in these nitrides is not related to a softening of the
phonon modes.
For B2 MX, a high density of states at the Fermi level may be responsible for the dynamical
instability of these phases.

\section*{Acknowledgements}
This work was supported by the STCU Contract No. 5539.
The work of P.\,T. was performed under the auspices of the U.\,S. Department of Energy by the Lawrence
Livermore National Laboratory under contract No.~DE--AC52--07NA27344.

\ukrainianpart

\title{Фазовий перехід з B1 в B2 у карбідах і нітридах титану та цирконію під тиском}
\author{В.І. Іващенко\refaddr{label1}, П.Е.А. Турчі\refaddr{label2}, В.І. Шевченко\refaddr{label1}}
\addresses{
\addr{label1} Інститут проблем матеріалознавства НАН України, вул. Кржижанівського, 3,
03142 Київ, Україна
\addr{label2} Лоуренсівська національна лабораторія в Ліверморі (Л--352),
поштова скринька 808, \\Лівермор, CA 94551, США
}

\makeukrtitle

\begin{abstract}
\tolerance=3000%
На основі першопринципних розрахунків електронної та фононної структур досліджувалась стабільність
різних фаз МХ (M\,=\,Ti, Zr; X\,=\,C, N) в рівновазі та під тиском.
Результати вказують на те, що всі B1 (типу NaCl) структури МХ при високому тиску зазнають
фазового перетворення в B2-структури, що узгоджується з попередніми розрахунками повної енергії.
Структури B1-MX динамічно стабільні при дуже високому тиску ($210\div570$~ГПа).
Обумовлені тиском B2 (типу CsCl) MC структури динамічно нестабільні навіть при високих тисках,
а TiN і ZrN кристалізуються в B2-структурі лише при тиску вищому $55$~ГПа.
Фазовий перехід першого порядку з B1 в B2 в цих нітридах не пов'язаний із пом'якшенням фононних
мод, а динамічна нестабільність B2-MX пов'язується з високою щільністю станів на рівні Фермі.
\keywords карбіди та нітриди титану і цирконію, першопринципні розрахунки, фазовий перехід,
електронні та фононні структури
\end{abstract}

\end{document}